# Deep Convolutional Neural Network for Roadway Incident Surveillance Using Audio Data

Zubayer Islam, Mohamed Abdel-Aty

*Abstract*—Crash events identification and prediction plays a vital role in understanding safety conditions for transportation systems. While existing systems use traffic parameters correlated with crash data to classify and train these models, we propose the use of a novel sensory unit that can also accurately identify crash events: microphone. Audio events can be collected and analyzed to classify events such as crash. In this paper, we have demonstrated the use of a deep Convolutional Neural Network (CNN) for road event classification. Important audio parameters such as Mel Frequency Cepstral Coefficients (MFCC), log Mel-filterbank energy spectrum and Fourier Spectrum were used as feature set. Additionally, the dataset was augmented with more sample data by the use of audio augmentation techniques such as time and pitch shifting. Together with the feature extraction this data augmentation can achieve reasonable accuracy. Four events such as crash, tire skid, horn and siren sounds can be accurately identified giving indication of a road hazard that can be useful for traffic operators or paramedics. The proposed methodology can reach accuracy up to 94%. Such audio systems can be implemented as a part of an Internet of Things (IoT) platform that can complement video-based sensors without complete coverage.

## I. INTRODUCTION

ROADWAY incident detection and reporting can prove to be an effective measure to reduce post-incident response time. In the United States, 38,680 people died in traffic crashes during 2020. In these accidents, 2 out of every 5 were alive when first responders arrived, but later died [1]. This clearly shows the need for real-time monitoring systems that can report accidents which can in turn lower deaths by improving post-crash response time. To this end, numerous studies have been proposed algorithms that can predict the incidents such as crashes based on real-time traffic parameters aggregated over 5-10 mins  [2-5] in Advanced Traffic Management and Information System (ATMIS). Traffic parameters has been correlated with crash incidents in most of these studies. While such system can identify and predict crashes with excellent accuracy, it is difficult to report other incidents such as hard brake resulting from tire skidding. Even multiple events of car horn at a particular location regularly can be an indicator of potential safety risk. In this paper, we propose a model that can identify tire skid sounds and car horns as well as crashes from audio data. Although environment sound classification has received increasing attention for noise surveillance and context awareness computing [6-8], it has not been considered in the transportation field from a safety point of view.

Crash events can be detected and predicted using traffic parameters such as speed, volume, etc. Extensive studies have been proposed in the past couple of decades showing the improvement in accuracy metrics across both statistical [9-11] and machine learning models [12-16]. Different types of data sources have been used in previous studies like video data [17], taxi data [18], probe vehicle GPS [19], weather [20], road network attributes, loop detector data [21], etc. A smart addition to the mix would be the integration of other sensor data that can identify certain hazardous roadway condition. Crash, tire skid and even car horn can be an indication of safety hazard which can be accurately classified using audio data.

As human beings, we are used to sensing a multitude of environmental sounds using hearing senses. Audio signals encompasses many scenes in the urban city scape. In roadways, we are alerted by the siren of a fire truck even though we may not see it around. Therefore, development of an intelligent system that can extract certain events using audio data has the potential to contribute to a smart city. Noise monitoring, security surveillance [22, 23], soundscape assessment [24], sound monitoring for improving hearing health [25], etc. are few applications that have used environment sound classification techniques. Roadway condition surveillance have been studied using audio data [26] but the study is limited to only two classes such as tire skid and crash. Therefore, it will not be able to differentiate between urban noises and roadway incidents. In this study we have included 17 different types of sounds that the model can learn from. Moreover, some studies have focused on in-vehicle sound detection [27]. Different signal processing and machine learning techniques have been used for classification of environmental sounds such as matrix factorization of spectral representations [28, 29], deep scattering [30], filter bank [31], etc. More recently, several deep learning architectures [32, 33] such as CNN, LSTM, etc. have been used. CNNs are of interest in audio classification since it is able to interpret spatio-temporal features from a 2D feature map. Moreover, the kernel filters that are used in CNN are able

Zubayer Islam (email: zubayer_islam@knights.ucf.edu) is a Postdoctoral Scholar and Mohamed Abdel-Aty (email: m.aty@ucf.edu) is a Pegasus Professor at the University of Central Florida, Orlando FL 32816.

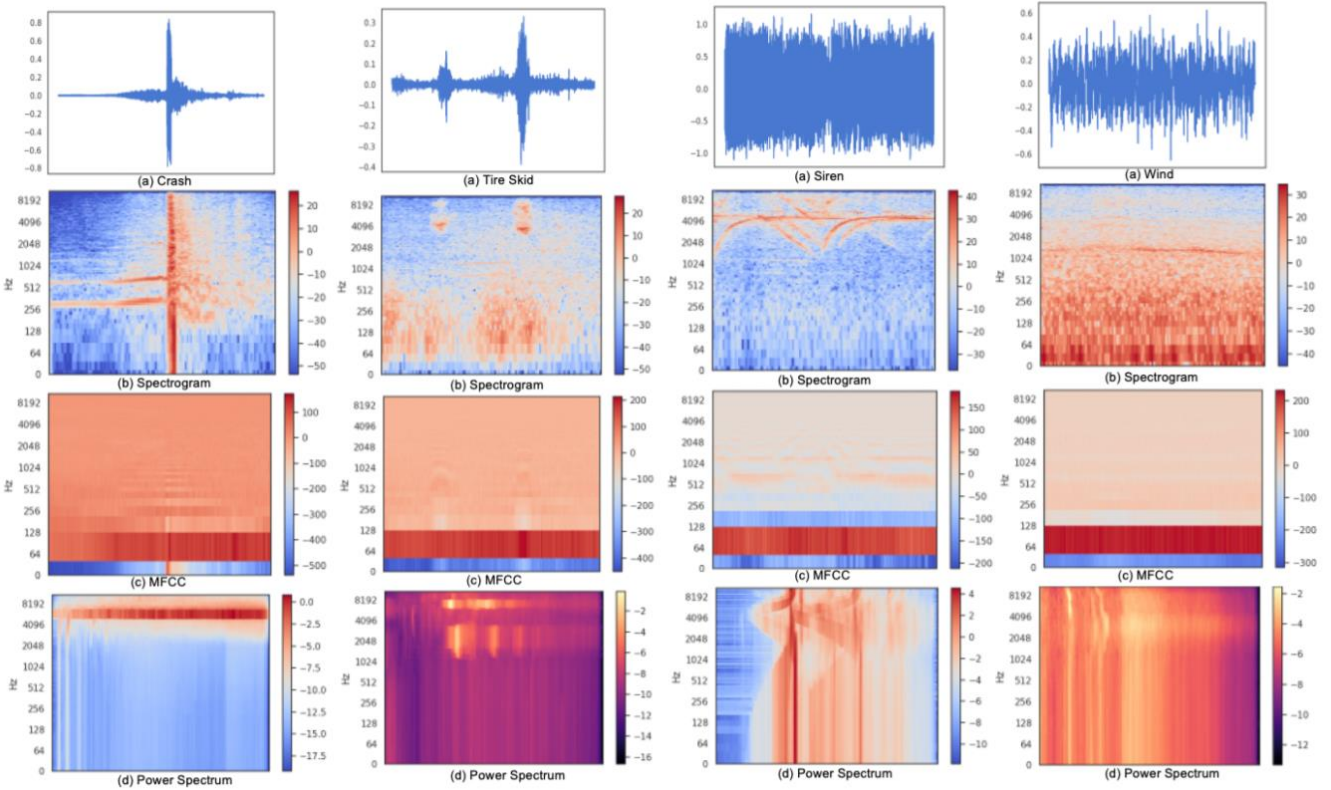

**Fig. 1.** Spectrogram, MFCC and Power spectrum of four different sound samples: crash, tire skid, siren, and wind

to learn representative features within a sound clip regardless of its position or length. Most studies in the literature use the spectrograms of audio clips to train CNN models yet there can be several advantages to the use of Mel Frequency Cepstral Coefficients (MFCC) and power spectral density. An elegant solution would be to use a feature set with these three 2D maps to form an RGB like image sample for audio classification. In this paper, we propose the use of a deep CNN architecture that can learn from these 3-channel features maps. A dataset containing crash and tire skid sounds was created by fusion with another open source dataset ESC-50 [34]. The minority samples were then augmented using time stretch, pitch shift, etc. The overall pipeline is able to reach an accuracy of 94%.

The main contributions of this paper can be highlighted as follows:
- Augmentation of existing systems with an added auditory sensor to improve safety. To the best of the knowledge of the authors this is the first comprehensive study that reports roadway incident detection using solely audio data.
- Use of a three-feature map for deep convolutional neural network in smart sound recognition (SSR).
- Use of data augmentation scheme to upsample the minority class

The next section describes the feature extraction process from the raw data. Section III highlights how the data was augmented, Section IV shows the model architecture followed by the experimental work section which describes the results from the different models.

## II. FEATURE EXTRACTION

Three different types of features were extracted from the audio sample data such as spectrogram images, Mel frequency cepstral coefficients (MFCC) and log Mel-filterbank power spectrum. The spectrogram images were generated by first converting the audio signals to their corresponding frequency spectrum using Short Term Fourier Transform (STFT). The spectrogram images show the frequency contents of the audio samples at any instant. This can be calculated using (1).

$$F(n,w) = \sum_{i=-\infty}^{i=\infty} x(i)w(n-i)e^{-jwn} \quad (1)$$

Here $x(i)$ is the input signal and $w(i)$ is the Hamming and Hanning window function. For the proposed dataset, we have sampled the audio at 5490 Hz followed by short-time Fourier transform (STFT) with windowed signal length of 860. These parameters were carefully chosen so that each of the three features had the exact dimension of (430,128). A sample of the extracted features is shown in Fig. 1.

Next the MFCC coefficients of the audio signal was estimated. It is one of the widely used features in speech recognition. Computing the MFCC comprises of five different steps such as dividing the signals into specified frames, obtaining the amplitude spectrum, taking the logarithm, converting to Mel-spectrum, and taking the discrete Fourier transform (DCT). Audio data can be converted into Mel spectrum using (2). We have used sampling rate of 44100 Hz with number of MFCC coefficients as 1000 to obtain the final

shape of (430, 120).

$$M(f) = 1125 \ln\left(1 + \frac{f}{700}\right) \quad (2)$$

The Mel frequency can relate the perceived frequency as heard by a human to its measured frequency. It is designed keeping in mind that humans can usually understand changes in pitch at low frequencies than at high frequencies. Therefore, the filters are narrower towards the low frequency components and wider near high frequencies as shown in Fig. 2. Using this feature ensures that the model is trained on values that humans can interpret. This is important since the ESC-50 dataset as well as the two classes added by the authors were manually labelled.

Thirdly, the power spectrum of the audio clips was also extracted. It can be given by the equation

$$E = \int_{-\infty}^{\infty} |x(t)|^2 dt \quad (3)$$

Power spectrum or power spectral density refers to the energy distribution of a time series signal x(t) into frequency components of the same signal. In other words, power spectrum describes how the energy of a signal changes with frequency. For the proposed dataset, we have computed the log Mel-filterbank energy features. The audio data was sampled at 22100 Hz. Number of filters was set to 128 with a window length of 0.71s. Several python libraries were used to obtain all the three different features [35, 36].

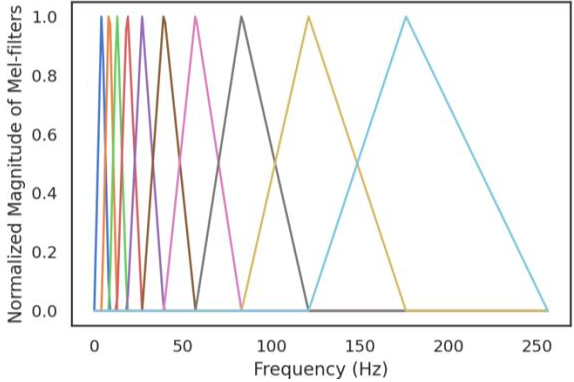

**Fig. 2.** Mel Filter

The raw audio samples were sampled at different rates so that the extracted 2-dimentional feature set were identical in shape. Four audio samples and the respective spectrogram, MFCC and power spectrum are shown in Fig. 1. Visual inspection shows that there are unique features in each of the classes that can be exploited by using a deep-CNN model.

### III. Data Augmentation

We have implemented four different types of audio deformation to the original data. Data augmentation aids in balancing the dataset. Since there are four minority classes in the dataset, the augmentation step helped to create representative samples from the minority set. The types of augmentations are briefly detailed below:

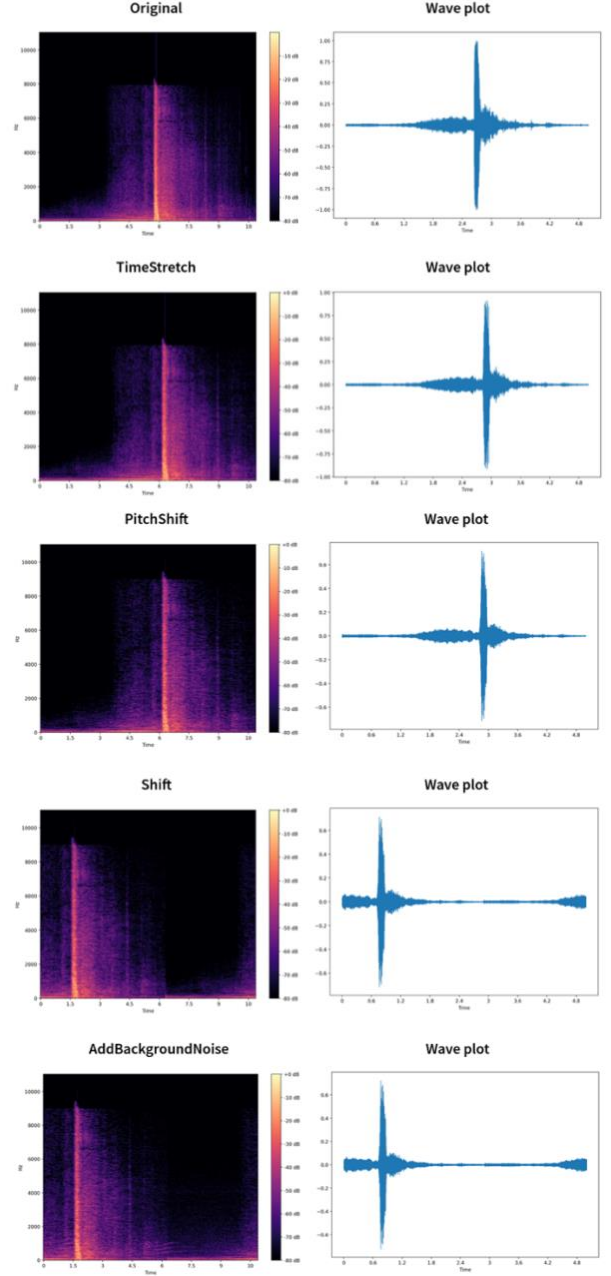

**Fig. 3.** Sample output of data augmentation

1. Background Noise: The audio sample was mixed with background noise from different scenario. The minimum and maximum amplitude was set to 0.001 and 0.015, respectively.
2. Time Stretch: A sound sample can be sped up or slowed using this feature without changing the pitch. A minimum and maximum rate was set at. 0.8 and 1.25, respectively. The function would randomly select a value from within this range.
3. Pitch Shift 1: The pitch of the audio wave was altered using this augmentation. Minimum and maximum

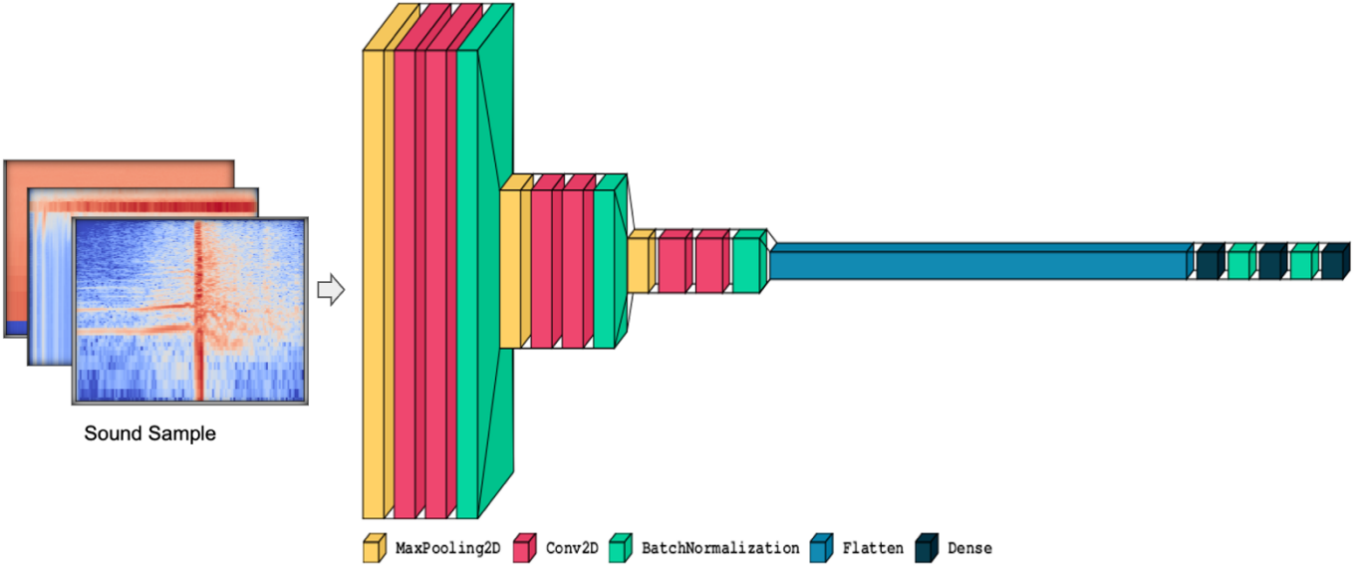

**Fig. 4.** Proposed Model Architecture

semitones were set as – 4 and 4 from which a value was randomly selected. Two more samples were generated by selecting slightly different semitones.

4. Shift: The audio sample was shifted in time using this technique. The minimum and maximum shift was set to half the audio length.

All the data augmentation was carried out using a Audiomentations [37], a python library. A few samples output is shown in Fig. 3.

## IV. MODEL ARCHITECTURE

A CNN model usually consists of various layers such as convolution, pooling, normalization, and fully connected layers. Sequential construction of the various layers aid the model to extract local features from input samples. CNNs were built from MLPs which are fully connected to one another. The specialty of CNN lies in the kernel filters F that is able to learn from spatially small local regions of the preceding layers. Moreover, it can extract high dimensional data and utilize complex architectures.

The convolution of feature map I with filter F can be given by (4)

$$(I * F)_{m,n} = \sum_{k=-a_1}^{a} 1 \sum_{l=-a_2}^{a} 2\, F_{k,l} I_{n-k,m-l} \quad (4)$$

These are followed by dense layers that are fully connected. Each layer also has an activation function. Two activation functions were used in the proposed model: rectified linear unit (ReLU) and softmax as given by equation (5) and (6).

$$g(z) = \max(0, z) \quad (5)$$

$$\sigma(x) = \frac{e^{x_i}}{\sum_{j=1}^{K} e^{x_j}} \quad (6)$$

The deep CNN architecture proposed in this work consists of six convolutional layers with max pooling layers in between. It is followed by the dense layers. The proposed model is shown in Fig. 4. The proposed model has the following layers which are parameterized as follows:

- Layer 1: 64 filter with a kernel size or filter size of (3,3). It is zero padded and also contains the activation function ReLU. It is preceded by a MaxPool layer. This helped in reducing the training time significantly than using the entire feature array without compromising the accuracy.
- Layer 2: 64 filters with a kernel size of (3,3) and same padding with ReLU activation. It is followed by a max pooling layer with a pool size of (3,3). This layer is followed by a batch normalization layer which helps in faster convergence for the proposed model.
- Layer 3: 128 filters with a receptive field of (3,3). This is followed by ReLU but without any pooling layer.
- Layer 4: 128 filters with a receptive field of (3,3). This is followed by ReLU and a pooling layer of shape (3,3). It is also followed by a bath normalization layer.
- Layer 5: 256 filters with a kernel size of (3,3) and same padding with ReLU activation.
- Layer 6: 256 filters with a kernel size of (3,3) and same padding with ReLU activation and followed by batch normalization.
- Layer 7: 80 hidden neurons with ReLU activation. It is preceded by the flatten layer.
- Layer 8: 40 hidden neurons with ReLU activation.
- Layer 9: 5 output layers with softmax as the activation function.

The overall process flowchart for data preparation and model training is shown in Fig. 5. First the data was padded with necessary zeroes to bring all the sound data to a uniform 5 second sample. Next the data was split into train-test samples before augmentation. This was necessary so that the model does not see some of the original data during the evaluation process. Thirdly, the data was augmented since the dataset is imbalanced. Six samples were generated from the original samples to increase the sample size seven-fold. From all the samples, spectrogram, MFCC and power spectrum images were extracted. A 3-channel image like data sample was generated afterwards and passed through the deep CNN model for training.

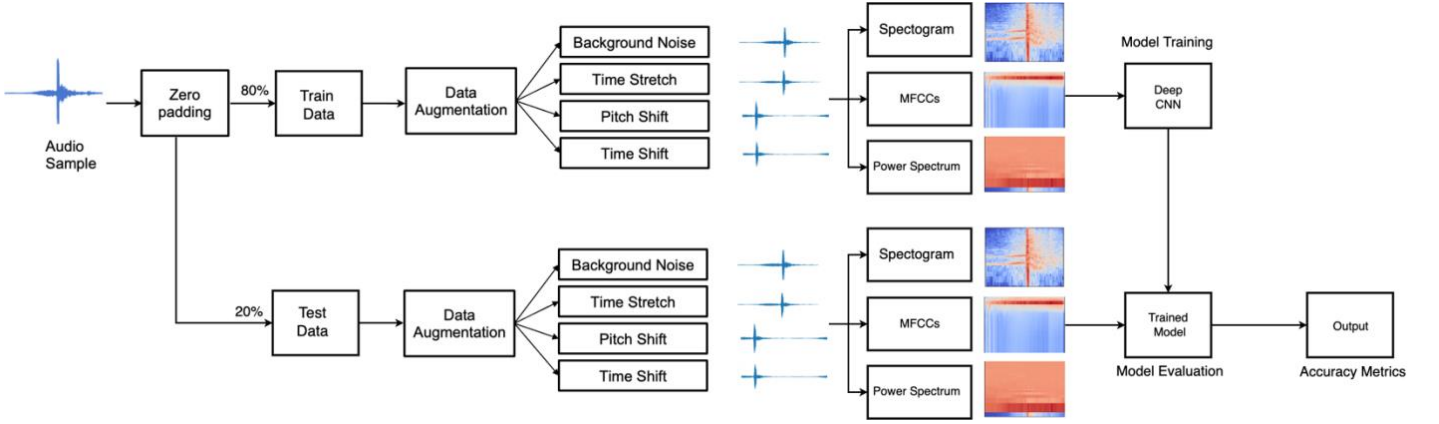

**Fig. 5** Model Implementation Flowchart

## V. EXPERIMENTAL WORKS

### A. Dataset

While there are several open-source environmental sound datasets [34, 38-40], special audio events such as crash, tire skidding events are not available. For this paper, the authors have expanded the existing ESC-50 sound dataset with crash and tire skid audio clips. These events were collected from different sources such as FreeSound [41], Zapslat [42], etc. Forty samples of crash and tire skid clips each were collected to balance the number of samples per class in the ESC-50 dataset.

TABLE I
DATASET DESCRIPTION

| Classes | Sound Type | Total Samples | Samples after augmentation | Source |
|---|---|---|---|---|
| Class 1: Urban Sound | Engine, Train, Helicopter, Airplane, Fireworks, Crying baby, Sneezing, Clapping, Coughing, Footsteps, Laughing, Rain, Wind | 520 | 520 | ESC-50 [34] |
| Class 2 | Crash Sound | 40 | 280 | [41, 42] |
| Class 3 | Siren | 40 | 280 | [34] |
| Class 4 | Tire Skid | 40 | 280 | [41, 42] |
| Class 5 | Car Horn | 40 | 280 | [34] |

The sound clips were manually labelled. The ESC-50 dataset has 2000 sound samples from 50 different classes. The average clip duration is around 5 seconds. All classes from the ESC-50 dataset were not used. The authors selected the classes based on the clips that can be usually heard on roadside. The selected sounds and related information are shown in Table I. It also shows the five different classes that the dataset was divided into. The number of samples in the minority class was increased seven folds by the use of the augmentation techniques.

### B. Evaluation Criteria

For evaluating the model performance on the engineered dataset, several key metrics were used such as accuracy, specificity, sensitivity, precision, and f1-score. The dataset was split into 80-20 ratio as training and test datasets. The performance metrics are reported for the test data. While accuracy is usually studied for balanced dataset, we need to check other parameters such as recall, precision, etc. as shown in (7-11). to get the exact model performance. Precision indicates what percentage of the predictions are relevant, recall indicated what percentage of relevant predictions are made while the false positive rate indicates how often the model is likely miss-predict. F1-score is calculated based on precision and recall.

$$Accuracy = \frac{TP + TN}{TP + TN + FP + FN} \quad (7)$$

$$False\ Positive\ Rate = \frac{FP}{TN + FP} \quad (8)$$

$$Recall = \frac{TP}{TP + FN} \quad (9)$$

$$Precision = \frac{TP}{TP + FP} \quad (10)$$

$$F1 - score = 2\frac{Precision * Recall}{Precision + Recall} \quad (11)$$

### C. Results

The results of the final model are shown in Table II. The overall accuracy of the model is 94% with average precision, recall and f1-score of individual classes as shown in Table II. The results indicate overall a very good performance in classifying audio events. The false positive rate is also low at 5.8%. The f1-score falls within the range 0.92 to 0.99 which indicates good balance between precision and recall.

TABLE II
MODEL RESULTS

| Class | Precision | Recall | F1-score | False Positive Rate | Total test samples |
|---|---|---|---|---|---|
| Urban Sound | 0.94 | 0.92 | 0.93 | 0.08 | 78 |
| Crash | 0.94 | 0.98 | 0.96 | 0.02 | 64 |
| Siren | 0.95 | 0.89 | 0.92 | 0.10 | 64 |
| Car Horn | 0.98 | 1.00 | 0.99 | 0.00 | 64 |
| Tire Skid | 0.89 | 0.91 | 0.90 | 0.09 | 64 |

The confusion matrix is also shown in Fig. 6. There were 64 samples in the test data for each of the events such as crash, siren, car horn and tire skid. The model could successfully classify over 57 of the samples per class. Moreover, it is also able to accurately classify the urban sounds to reduce the false alarms. One percent of the crashes were not identified correctly and 9% of the tire skid sounds were also missed. The overall false positive rate of the model is about 6%. The confusion matrix also shows that 7 urban sound classes were also reported as incident samples which is only 9% of the total samples in urban sound. Therefore, we can say that the model performs reasonably well to classify special road events as well as filters out common urban noises.

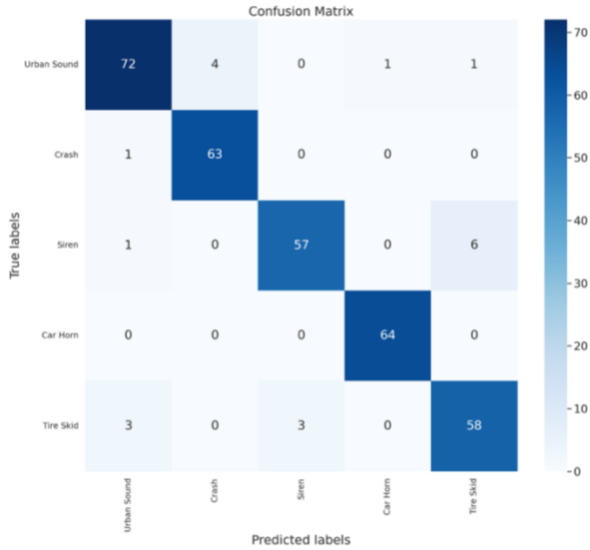

**Fig. 5.** Confusion Matrix

The proposed model was able to differentiate between the five different types of classes as can be seen in Fig. 7. t-SNE (t-Distributed Stochastic Neighbor Embedding) [43] was used to encode the high dimensional data for visualization. It can reduce multiple features into two or three values for observation. The Fig. 7(a) shows the feature values from the first convolutional layer in the model while Fig. 7(b) shows those from the final dense layer. It is evident that the model is successfully able to distinguish between the different types of sounds.

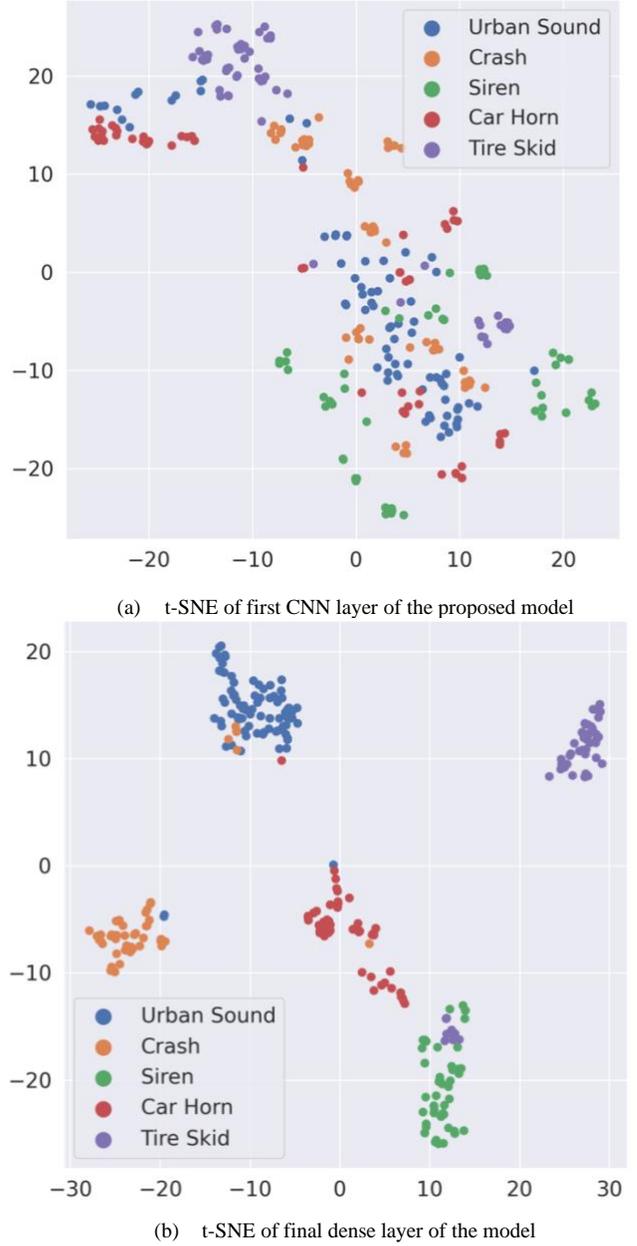

(a) t-SNE of first CNN layer of the proposed model

(b) t-SNE of final dense layer of the model

**Fig. 7.** t-SNE

*D. Model Comparison*

The proposed model has been compared to several benchmark machine learning models such as Random Forest (RF), k-Nearest Neighbors (KNN), Support Vector Machine (SVM), XGBoost (XGB) and LightGBM. RF, KNN, SVM, XGB, LightGBM were trained on the original dataset without any augmentation. This helps to understand whether the augmentation is useful or not. The proposed CNN model was trained on two different datasets. Both the CNN models were trained on the augmented dataset. CNN-1 was trained on only the MFCC feature values while CNN-3 was trained on all the three feature sets (Spectrogram, MFCC and Power Density). Comparing the two CNN models, it can be interpreted whether the three-channel feature set is providing better results. The results are shown in Fig. 8 in terms of recall, precision, f1-score and accuracy. The boxplots were generated using a 10-fold cross validation where 30% data was set aside for evaluation in each fold.

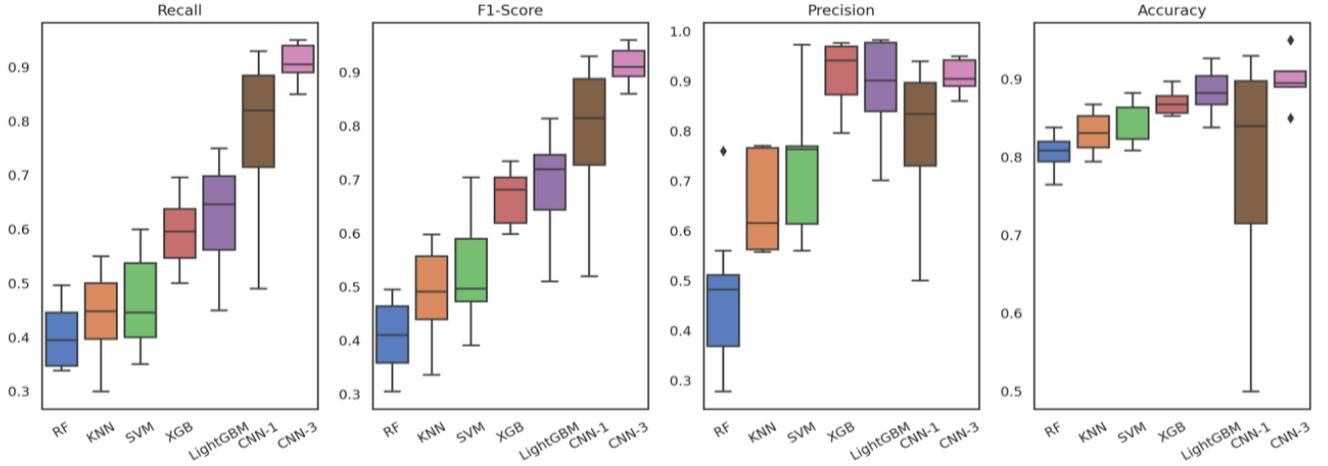

**Fig. 8.** Comparison between proposed model and benchmark models in terms of recall, f1-score, precision, and accuracy

The CNN-3 model produces better results than all other models in terms of recall and f1-score. For the dataset shown in this study, recall is the most important metric since it shows how accurately the model can identify the critical events such as crash, tire skid, etc. CNN-1 also gives better accuracy than the non-augmented dataset showing that augmentation improved performance. The downside of the CNN-1 model is that it had quite some large variation in the all the metrics. CNN-3 outperforms CNN-1 is an indicator that the three-channel feature set was also useful for training the model.

It can also be noted that the precision of some of the models is better than CNN 3 for example XGB and LightGBM. since these models well trained on the original data set, it is likely that the models predicted the majority class more accurately than the minority samples. Thus, the precision is higher, but the recall is low. From the model accuracy we also see that CNN 3 model is the best performing model with lower variance. Overall comparing the four different metrics it is evident that CNN-3 model performs significantly better and all the other benchmark models.

The proposed model is also compared to several studies from the literature as shown in Table III. Most of the studies have a limited number of classes in the data set which makes it inconvenient to differentiate between hazardous and non-hazardous events. Moreover, none of these studies have taken the advantage of audio data augmentation. Hence the proposed methodology as well as the data set yields a better performance than all the existing studies.

TABLE III
COMPARISON WITH STUDIES FROM THE LITERATURE

| Study | Accuracy | No of Classes | Model | Parameters |
|---|---|---|---|---|
| Foggia et. al [26] | 0.71 | 2 | SVM | Dictionary Learning |
| Sammarco et. al [27] | 0.88 | 11 | Random Forest | Time and Frequency Series values |
| Mnasri et al [44] | 0.90 | 3 | MLP, LSTM, BLSTM | Energy, Power, Spectral and Temporal features |
| Li et. al [45] | 0.92 | 2 | BLSTM | MFCC, Filterbank |
| Proposed Model | **0.94** | **17** | Deep CNN | MFCC, Spectrogram, Power Spectrum |

## VI. CONCLUSION

This study proposes a deep convolutional neural network architecture than can differentiate between the hazardous road incidents such as crash, tire skid, etc. and common urban sounds. The audio data was collected from various sources and combined with an open-source dataset. It was further augmented to create more representative samples from the minor classes. Three different features were extracted from all the samples and processed to form a three-channel image like input for the deep CNN architecture. The results show good overall metrics in terms of accuracy, recall and f1-score. The system can be implemented in roadway networks that have audio and video recording capabilities. As a result of the roadway incident notifications emergency services response time can be improved thereby saving lives. Moreover, repeated roadway incidents such as tire skid or car horn at a particular location can be indicative of certain safety hazard. Future studies can investigate the safety of roadways using audio data features. Microphones are usually cheap and easy to install. Audio detection could extend the detection capability to remote areas where CCTVs are not common, or work in conjunction with other sensors to achieve better coverage.